\documentclass[journal]{IEEEtran}

\usepackage{amsmath,amssymb}
\usepackage{graphicx}
\usepackage{booktabs}
\usepackage{cite}
\usepackage{siunitx}
\usepackage{comment}
\usepackage[colorlinks=true,linkcolor=black,citecolor=black,urlcolor=blue]{hyperref}

\newcommand{\EsN}{E_s/N_0}
\newcommand{\R}{\mathcal{R}}
\DeclareSIUnit{\symbol}{symbol}
\DeclareSIUnit{\bps}{bps}
\begin{document}

\title{FTN Signaling: Spectral Efficiency from BPSK to 16-QAM}

\author{Emre~Cerci,~Adem~Cicek,~Melda~Yuksel,~\IEEEmembership{Senior~Member,~IEEE},~Gokhan~M.~Guvensen,~\IEEEmembership{Member,~IEEE},\\~Enver~Cavus,~Yaser~Dalveren,~and~Halim~Yanikomeroglu,~\IEEEmembership{Fellow,~IEEE}

\thanks{Manuscript received Month XX, 2026. This work was supported by the Scientific and Technological Research Council of T\"urkiye (T\"UB\.ITAK) under Grant 122E236.}
\thanks{E. Cerci is with the Department of Electrical and Electronics Engineering, Atilim University, 06830 Ankara, T\"urkiye (e-mail: emre.cerci@atilim.edu.tr).}
\thanks{A. Cicek is with the Department of Electrical and Electronics Engineering, Cankiri Karatekin University, 18200 Cankiri, T\"urkiye (e-mail: ademcicek@karatekin.edu.tr).}
\thanks{M. Yuksel and G. M. Guvensen are with the Department of Electrical and Electronics Engineering, Middle East Technical University, 06800 Ankara, T\"urkiye (e-mail: ymelda@metu.edu.tr; guvensen@metu.edu.tr).}
\thanks{E. Cavus is with the Department of Electrical and Electronics Engineering, Ankara Yildirim Beyazit University, 06010 Ankara, T\"urkiye (e-mail: ecavus@aybu.edu.tr).}
\thanks{Y. Dalveren is with the Department of Electrical and Electronics Engineering, Izmir Bakircay University, 35665 Izmir, T\"urkiye (e-mail: yaser.dalveren@bakircay.edu.tr).}
\thanks{H. Yanikomeroglu is with the Department of Systems and Computer Engineering, Carleton University, Ottawa, ON K1S 5B6, Canada (e-mail: halim@sce.carleton.ca).}}

\markboth{IEEE Communications Letters, Vol.~XX, No.~X, Month~2026}%
{Cerci \MakeLowercase{\textit{et al.}}: FTN Signaling: Spectral Efficiency from BPSK to 16-QAM}

\maketitle

\begin{abstract}

Faster-than-Nyquist (FTN) signaling can improve spectral efficiency by transmitting symbols closer together than the Nyquist limit, at the cost of additional inter-symbol interference (ISI). In this letter, we study the finite-alphabet symbol-wise achievable information rate (AIR) of FTN signaling for BPSK, QPSK, and 16-QAM under a transmit-power constraint and a common minimum mean square error (MMSE) channel-shortening (CS) detection framework, cross-validated for BPSK against an independent reduced-state Ungerboeck Bahl-Cocke-Jelinek-Raviv (BCJR) benchmark. Our results show a clear trend. Lower-order constellations can benefit from more aggressive time acceleration, while higher-order constellations become less tolerant of acceleration and achieve their best spectral efficiency closer to the Nyquist limit. At a reference signal-to-noise ratio (SNR) of $\SI{6}{\decibel}$ the optimum acceleration factor shifts from $\tau^\star\approx0.65$ for BPSK to $0.75$ for QPSK and $0.90$ for 16-QAM. We further show that this trend is preserved in coded systems and remains robust to the detector memory. The impact of FTN operation on the instantaneous-to-average power ratio (IAPR) is also examined. Overall, the results highlight how constellation order should be considered when selecting the operating point for FTN signaling.
\end{abstract}


\IEEEpeerreviewmaketitle

\section{Introduction}
\IEEEPARstart{F}{aster}-than-Nyquist (FTN) signaling transmits one symbol every $\tau T$ seconds, with $0<\tau\le1$, and thus packs $1/\tau$ times more symbols into the same bandwidth at the cost of inter-symbol interference (ISI) \cite{mazo1975ftn,anderson2013ftn}. A natural way to evaluate this trade-off between rate and ISI is through the achievable information rate (AIR) \cite{arnold2006air}, measured in $\unit{\bit\per\symbol}$, expressed as spectral efficiency (SE) in $\unit{\bit\per\second\per\hertz}$.

The information-theoretic limits of FTN signaling with Gaussian inputs are well understood \cite{rusek2009constrained,ganji2020capacity,zhang2025mimo}, and transmitter-side processing can reduce the resulting ISI through precoding \cite{jana2017preeq,wang2017linear,wang2020iterative}. Finite-alphabet information rates have also been studied, including the constrained capacities of Rusek and Anderson \cite{rusek2009constrained}, information rates for binary, quaternary, and octal signaling \cite{rusek2006rates}, and achievable rates with reduced-complexity FTN receivers \cite{prlja2012reduced}. Most of these studies, however, focus on a particular modulation format or operating point. Here, we look more broadly at how the SE-maximizing acceleration factor changes with constellation order, using a common constant-power constraint and the same detection framework. We then interpret the observed differences in terms of the minimum Euclidean distance of the constellations.

In this letter, we compute the finite-alphabet symbol-wise AIR for BPSK, QPSK, and 16-QAM under a fixed power normalization, and \emph{the same} $M$-ary channel-shortening (CS) minimum mean square error (MMSE) detector. A reduced-state Ungerboeck Bahl-Cocke-Jelinek-Raviv (BCJR) detector serves only as an independent, near maximum a posteriori probability (MAP) benchmark to validate the CS-MMSE rate for BPSK.

The main finding is that all three constellations benefit from FTN acceleration. For each one, there is a $\tau^\star<1$ at which the SE exceeds its Nyquist ($\tau=1$) value. Interestingly, however, the benefit becomes smaller as the constellation order increases, while the corresponding $\tau^\star$ moves closer to one. We further (i) verify with a larger-memory ($\nu=2$) 16-QAM detector that the observed optimum is not caused by detector limitations, (ii) quantify the bit-wise generalized mutual information (GMI) penalty relevant to coded BICM systems, and (iii) examine the instantaneous-to-average power ratio (IAPR) cost of FTN acceleration at these operating points.

\section{System Model}
With the unit-energy root-raised cosine (RRC) pulse $h(t)$ and unit-\emph{average}-energy symbols $a_n$ from an $M$-ary constellation, the transmitted signal is $s(t)=\sqrt{E_s}\sum_n a_n h(t-n\tau T)$. After matched filtering and symbol-rate sampling the standard FTN model is $z_n=\sum_m g_m\,a_{n-m}+\eta_n,$
\noindent where the ISI taps are the samples of the pulse autocorrelation,
\begin{equation}
    g_m=\int_{-\infty}^{\infty} h(t)h(t-m\tau T)dt,
    \label{eq:taps}
\end{equation}
\noindent with $g_0=1$, $g_{-m}=g_m$, and \emph{colored} noise of covariance $\mathbb{E}[\eta_n\eta_{n'}^{*}]=N_0\,g_{n-n'}$. A soft-output detector returns the posterior $q= P(a_n|\mathbf z)$, and the per-symbol AIR is the symbol-wise mismatched-decoding lower bound
\begin{equation}
\R=k+\mathbb{E}\!\left[\log_2 q
\!\left(a_n=a_n^{\text{true}}|\mathbf z\right)\right], \qquad k=\log_2 M, \label{eq:air}
\end{equation}

\noindent where $a_n^{\text{true}}$ is the symbol actually transmitted at time $n$ and $q$ is the posterior computed by the detector. Because $q$ generally differs from the true posterior, this score can only \emph{underestimate} the information the channel conveys. As a result, \eqref{eq:air} is an achievable \emph{symbol-wise} rate lower bound for \emph{any} detector, i.e., for mismatched decoding. On the other hand, it equals the true symbol-wise mutual information when $q$ matches the channel. The SE follows as
\begin{equation}
\mathrm{SE}=\frac{\R}{\tau\,(1+\beta)}\quad[\text{bits/s/Hz}]. \label{eq:se}
\end{equation} 

We hold the total transmit power constant when sweeping $\tau$. Therefore, the per-symbol SNR entering the detector is
\begin{equation}
\EsN=\rho_1\tau, \label{eq:power}
\end{equation}
where $\rho_1$ is the reference signal-to-noise ratio (SNR) at $\tau=1$. This normalization keeps the comparison fair. If one instead held $\EsN$ fixed while decreasing $\tau$, the total transmit power would grow proportionally to $1/\tau$ and the SE would increase monotonically with acceleration, hiding the trade-off studied here. 

For reference, we also report the upper bound obtained with independent complex-Gaussian input symbols under the same power constraint, in bit per symbol,
\begin{align}
C_g&=\int_{0}^{1}\log_2\!\bigl(1+(\EsN)\,X(f)\bigr)\,df,
    \label{eq:cg}
\end{align}
\noindent where $X(f)=\sum_{m=-\infty}^{\infty} g_m e^{-j2\pi fm}$ is the \emph{folded spectrum} of the ISI taps $g_m$ \cite{ganji2020capacity}, normalized so that $\int_{0}^{1}X(f)\,df=g_0=1$.

\section{Estimation Methodology}
This section describes the common detector, justifies its short memory for 16-QAM, and lists the validation checks applied to every reported rate.
\subsection{Unified $M$-ary CS-MMSE Detector}
\label{sec:detector}
The common detector for \emph{all} constellations is a CS-MMSE detector \cite{rusek2012channel}. The MMSE prefilter is expressed in the frequency domain as $W(f)=E_s/(E_sX(f)+N_0)$. This filter reshapes the long FTN channel into a linear system with an effective impulse response whose $2\nu{+}1$ dominant central taps are retained as a short target. Here, $\nu$ denotes the one-sided span of this shortened target. The remaining taps and filtered noise are modeled as additional white Gaussian noise of matched variance, and the shortened response is applied as a \emph{causal} channel, so the working detector is an exact Forney-type BCJR over its full $2\nu$-symbol memory, with $M^{2\nu}$ trellis states in our implementation.

This causal implementation is conservative in complexity. An Ungerboeck-type shortening receiver as in \cite{rusek2012channel} would attain $M^{\nu}$ states for the same target span. This trades a small, quantifiable model mismatch for an exponentially smaller trellis. The target memory $\nu$ is increased until the rate saturates: QPSK uses $\nu=3$ ($4096$ states) and 16-QAM $\nu=1$ ($256$ states), with the adequacy of $\nu=1$ for 16-QAM established in the next subsection.

\subsection{16-QAM Complexity and $\nu$-Convergence}
\label{sec:16qam}
The question here is whether the short memory $\nu=1$ is sufficient for 16-QAM, i.e., whether the reported optimum reflects the information rate itself rather than a receiver limitation. Reduced-state hard pruning is not a remedy---in deep ISI it can drop the true path and produce invalid negative auxiliary rates, which are discarded. We therefore use the $\nu=1$ detector for the main sweep and certify its adequacy in two ways.

Let $\mathbf{c}$ denote the full response after the prefilter and $\mathbf{b}$ its $2\nu{+}1$ retained central taps. The first check is the captured-energy fraction $E_{\text{cap}}=\lVert\mathbf b\rVert^2/\lVert\mathbf c\rVert^2$, the share of the effective response energy that the short target represents. We adopt $E_{\text{cap}}\ge0.99$ as a self-consistency criterion: it is not a proof of MAP optimality, but it guarantees that the mismatch between the assumed and the actual channel is small. For $\beta=0.3$, this criterion is met for $\tau\geq0.85$, which we call the reliable window. The results for 16-QAM outside this window are reported as looser lower bounds.

Next, to check directly whether the optimum is merely where the constrained receiver falls, we implemented a fast vectorized CS-MMSE BCJR, validated to reproduce the reference detector to within $10^{-9}$ bit at $\nu\leq1$, and ran the \emph{full} $\nu=2$ ($65{,}536$-state) detector at $\tau\in[0.80,1.00]$ as a reference for $\nu=1$ results.

\subsection{Validation}
\label{sec:validation}
Since \eqref{eq:air} is only a lower bound, three checks guard the tightness and the bias of the estimates. First, every estimator must pass an interference-free \emph{gate}: at $\tau=1$ the AIR must equal the known additive white Gaussian noise (AWGN) symbol-wise mutual information. For all three constellations the Monte-Carlo estimate at $\tau=1$ reproduces the analytical AWGN mutual information to within $0.005$~bit, so the estimators carry no systematic bias, and the AIR stays below $C_g$ for every simulated $\tau\in[0.5,1]$. Second, because a lower bound is most useful when it is known to be nearly tight, we compare against a near-optimal reference where one is still affordable. For BPSK we evaluate \eqref{eq:air} with a BCJR \cite{bahl1974optimal} on the \emph{Ungerboeck} (matched-filter, colored-noise) metric \cite{ungerboeck1974adaptive}, which needs no whitening and therefore works below the realizability limit $\tau\ge1/(1+\beta)\approx0.77$ of spectral-factorization detectors \cite{jana2017preeq}. An $M$-algorithm \cite{prlja2012reduced} retains only the $M_{\text{red}}$ most likely of the $2^L$ trellis states, where $L$ is the retained ISI memory. At $\tau=0.6$ (again for $\beta=0.3$ with $L=12$), $M_{\text{red}}=128$ reproduces the AIR of the unpruned $4096$-state BCJR, giving a \emph{validated near-MAP} reference. The CS-MMSE BPSK rate matches this benchmark within the confidence interval at every $\tau$, justifying the use of the single CS-MMSE detector for the cross-constellation comparison. Third, shortcuts that fail these checks are rejected. A closed-form expectation-propagation rate is optimistic, and an iterative generalized approximate message passing (GAMP) rate is unstable for moderate ISI. In all estimates, the per-symbol term in \eqref{eq:air} may be \emph{negative} and is averaged as-is. Clamping it to zero would inflate the AIR. Each point is averaged over independent frames and reported with a $95\%$ confidence interval.

\section{Results and Discussion}
All results use an RRC pulse with roll-off $\beta=0.3$, constant transmit power, and the common CS-MMSE detector as explained in Section~III. Unless stated otherwise, $\rho_1=\SI{6}{\decibel}$.

\begin{figure}[t]
\centering
\includegraphics[width=0.96\columnwidth]{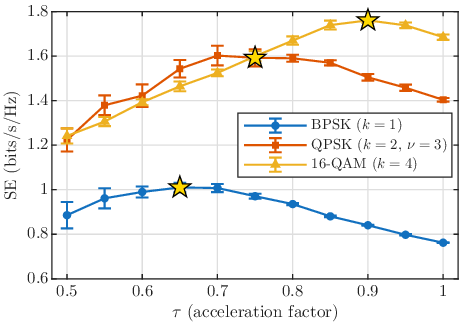}
\caption{SE vs.\ $\tau$ for BPSK, QPSK, and 16-QAM at $\rho_1=\SI{6}{\decibel}$ (stars: optima, error bars: $95\%$ confidence intervals).}
\label{fig:overlay}
\end{figure}

\subsection{Optimum SE for Different Constellations}
Fig.~\ref{fig:overlay} and Table~\ref{tab:summary} present the SE curves and their optima. All three curves rise to a peak at some $\tau^\star<1$ and then fall, but the SE gain over Nyquist drops from $33\%$ to $14\%$ to $4.5\%$ for BPSK, QPSK, and 16-QAM, while $\tau^\star$ moves from $0.65$ to $0.75$ to $0.90$.
For QPSK, the optimum lies on a flat plateau, $\tau^\star\in[0.70,0.80]$, indistinguishable within the confidence intervals, with the nominal peak at $0.75$.

Only $\nu=3$ results are reported for QPSK. A $\nu=2$ versus $\nu=3$ comparison on identical noise realizations showed that $\nu=2$ under-reports the rate by up to $\SI{0.015}{\bit}$ near the optimum while leaving $\tau^\star$ unchanged. The $\nu=2$ results are therefore omitted.

At $\tau=1$ there is no ISI, so the SE equals the AWGN symbol-wise AIR of the constellation at $\rho_1$ divided by $(1+\beta)$, and is set by the operating SNR rather than by the ceiling $k=\log_2M$. For 16-QAM at $\SI{6}{\decibel}$ this gives $2.20/(1+\beta)\!=\!\SI{1.69}{\bps\per\hertz}$ (the complex Gaussian capacity is $\SI{2.32}{\bit\per\symbol}$) rather than $4/(1+\beta)\!=\!\SI{3.08}{\bps\per\hertz}$. Its peak (Table~\ref{tab:summary}) lies inside the reliable window $\tau\ge0.85$ and its separation from the $\tau=1$ value exceeds the combined confidence intervals, so the modest gain is statistically significant. Points below $\tau=0.85$ are looser lower bounds.

For 16-QAM, the full $\nu=2$ detector of Section~III-B reproduces the $\nu=1$ SE within the confidence intervals ($|\Delta\R|\le0.003$~bit on identical noise realizations) and keeps the optimum at $\tau^\star=0.90$, so the optimum is information-limited rather than detector-limited.

\begin{table}[t]
\centering
\caption{SE optimum and gain over Nyquist for each constellation at $\rho_1=\SI{6}{\decibel}$.}
\label{tab:summary}
\setlength{\tabcolsep}{3.5pt}
\resizebox{\columnwidth}{!}{%
\begin{tabular}{lcccccc}
\toprule
                 & $\tau^\star$ & SE$(\tau{=}1)$ & peak SE             & gain vs. & reliable     & $\tau{=}1$\\
                 &              & [bps/Hz]       & [bps/Hz]            & Nyquist  & $\tau$ range & gate\\
\midrule
BPSK ($k{=}1$)   & 0.65         & 0.763          & $1.013\!\pm\!0.018$ & $32.8\%$ & 0.50--1.00   & PASS\\
QPSK ($k{=}2$)   & 0.70--0.80   & 1.404          & $1.602\!\pm\!0.044$ & $14.1\%$ & 0.50--1.00   & PASS\\
16-QAM ($k{=}4$) & 0.90         & 1.685          & $1.761\!\pm\!0.012$ & $4.5\%$  & 0.85--1.00   & PASS\\
\bottomrule
\end{tabular}}
\end{table}

\subsection{Minimum-Distance View of Constellation-Dependent $\tau$}
The order dependence seen in Fig.~\ref{fig:overlay} and Table~\ref{tab:summary} follows from the power constraint. Decreasing $\tau$ increases the symbol rate but reduces the per-symbol SNR according to \eqref{eq:power}. At unit average energy the constellations differ in their \emph{minimum Euclidean distance} $d_{\min}$, which is progressively smaller for QPSK and 16-QAM than for BPSK.
A smaller $d_{\min}$ makes a constellation more sensitive to the same SNR reduction, so its symbol-wise AIR falls more steeply as $\tau$ decreases. The optimum $\tau$ is the point where the $1/\tau$ packing gain balances this AIR loss, and the balance tips toward larger $\tau$ as the constellation order grows.

The pulse shape sets the \emph{structure} of the loss. For $\tau<1/(1+\beta)$ the folded spectrum $X(f)$ develops nulls, the Gaussian-input capacity saturates \cite{ganji2020capacity}, and the captured-energy fraction $E_{\text{cap}}$ of the shortening receiver collapses. Denser QAM grids, with their smaller decision regions, turn the same residual interference and SNR reduction into more frequent symbol errors, so their AIR degrades before the $1/\tau$ benefit in \eqref{eq:se} can be realized.
Consequently, low-order constellations can exploit aggressive acceleration even for $\tau<1/(1+\beta)$ ($\tau^\star=0.65$ for BPSK), whereas 16-QAM is confined to a mild $\tau^\star=0.90$.

\begin{figure}[t]
\centering
\includegraphics[width=0.92\columnwidth]{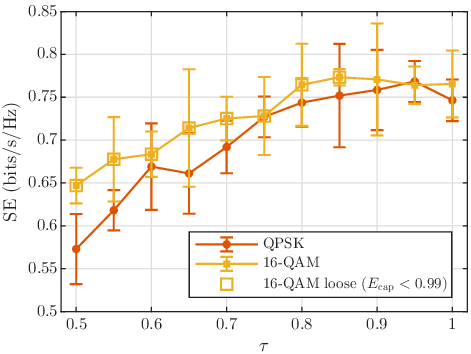}
\caption{SE vs.\ $\tau$ for QPSK and 16-QAM at $\rho_1=\SI{0}{\decibel}$ (open squares: 16-QAM points with $E_{\text{cap}}<0.99$).}
\label{fig:crossing}
\end{figure}

\subsection{Can a Lower-Order Constellation Overtake a Higher One?}
Fig.~\ref{fig:crossing} examines whether a lower-order constellation can overtake a higher one at low SNR, $\rho_1=\SI{0}{\decibel}$. Under constant power this is not excluded in principle, since the denser constellation suffers more from the SNR reduction and from residual ISI as $\tau$ shrinks.
In the measurements, however, the QPSK and 16-QAM SE curves do not cross. At $\rho_1=\SI{6}{\decibel}$ (Fig.~\ref{fig:overlay}) they only meet at the strongest simulated acceleration, $\tau=0.50$ ($1.223\pm0.052$ vs.\ $1.241\pm0.034\,\unit{\bps\per\hertz}$, inseparable within the confidence intervals), and the $\SI{0}{\decibel}$ sweep in Fig.~\ref{fig:crossing} shows no statistically significant QPSK~$>$~16-QAM region either. Since the 16-QAM points with $E_{\text{cap}}<0.99$ are looser lower bounds, an apparent crossing there would indicate a receiver limitation rather than a reordering, and no ranking conclusion is drawn in that region.

\subsection{Bit-wise GMI for Coded BICM}
\label{sec:gmi}
\begin{figure}[h!]
\centering
\includegraphics[width=.92\columnwidth]{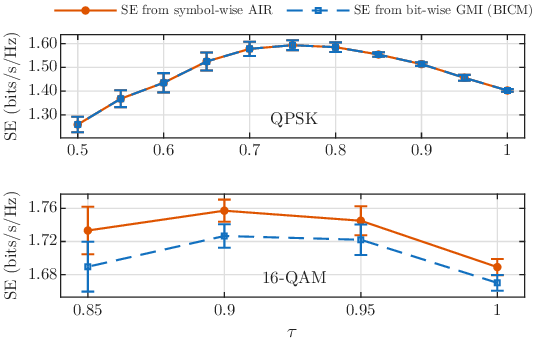}
\caption{SE vs.\ $\tau$ from the symbol-wise AIR (solid) and from the bit-wise GMI for BICM (dashed) for QPSK (top) and 16-QAM (bottom) at $\rho_1=\SI{6}{\decibel}$.}
\label{fig:gmi}
\end{figure}

Figs.~\ref{fig:overlay} and \ref{fig:crossing} report symbol-wise rates, whereas a coded receiver works with bit-wise metrics (BICM). Fig.~\ref{fig:gmi} therefore compares the SE obtained from the symbol-wise AIR with that obtained from the bit-wise GMI, $\mathrm{GMI}=\sum_{i=1}^{k}\big(1+\mathbb{E}[\log_2 q_i(b_i^{\text{true}}\mid\mathbf z)]\big)\le\R$, where $q_i(b\mid\mathbf z)$ are the per-bit posteriors of the same CS-MMSE detector under Gray labeling. For QPSK the two coincide at every $\tau$ (penalty $<0.1\%$), since the in-phase and quadrature bit sub-channels are separable. For 16-QAM the penalty is small, $1.1\%$ at $\tau=1$ and $1.7\%$ at the optimum, and $\tau^\star=0.90$ is preserved. The order-dependent trend thus survives under the coded-system metric.

\begin{figure}[t!]
\centering
\includegraphics[width=.92\columnwidth]{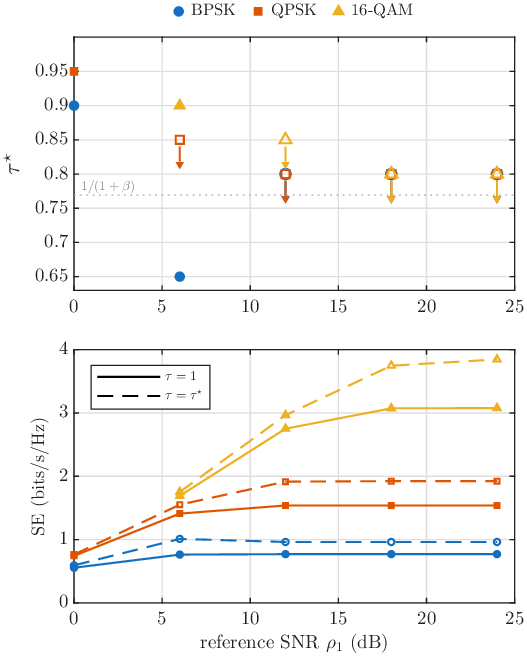}
\caption{Top: $\tau^\star$ vs.\ $\rho_1$ for BPSK, QPSK, and 16-QAM (filled: information-limited optima, open with arrows: detector-limited lower bounds). Bottom: SE vs.\ $\rho_1$ at $\tau=1$ (solid) and at $\tau=\tau^\star$ (dashed).}
\label{fig:oppoint}
\end{figure}

\subsection{Operating-Point Dependence}
As shown in Fig.~\ref{fig:oppoint} and Table~\ref{tab:power}, the optimum acceleration factor is also operating-point dependent. Repeating the $\tau$ sweeps at reference SNRs $\rho_1=0$--$\SI{24}{\decibel}$ shows that $\tau^\star$ moves to smaller values as $\rho_1$ increases. This behavior is expected: at high SNR, the finite-alphabet AIR approaches its ceiling of $k=\log_2 M$ bits per symbol, so the $1/\tau$ symbol-rate gain dominates over a wider range of $\tau$. The lower panel of Fig.~\ref{fig:oppoint} shows the SE at $\tau=1$ and at the gated $\tau^\star$. Both saturate toward the $k/(1+\beta)$ ceilings, and the FTN gain widens the gap between them. The receiver needs no structural change across operating points, only the operating SNR, from which the prefilter and the BCJR metrics are recomputed via \eqref{eq:power}. At high SNR the tabulated optima sit at the edge of the $E_{\text{cap}}$ reliability gate rather than at the optimum itself (marked with $^\dagger$ in Table~\ref{tab:power} and shown as open markers with downward arrows in Fig.~\ref{fig:oppoint}). They should be read as lower bounds $\tau^\star\le0.80$ on the useful acceleration depth. This is also why the dashed BPSK curve in Fig.~\ref{fig:oppoint} dips above $\SI{6}{\decibel}$. At $\rho_1\ge\SI{12}{\decibel}$ the gate confines the sweep to $\tau\ge0.80$, where the saturated AIR of $1$~bit gives $\mathrm{SE}=1/(0.8(1+\beta))=0.96$~bps/Hz, whereas at $\SI{6}{\decibel}$ the gate admits $\tau^\star=0.65$. Thus, the useful FTN acceleration should be selected jointly with the constellation order, operating SNR, and receiver complexity. The gains in Table~\ref{tab:power} come from a separate sweep with the $\nu=2$ QPSK detector and the $E_{\text{cap}}\ge0.99$ gate applied contiguously from $\tau=1$. They are therefore conservative lower bounds rather than the nominal optima of Table~\ref{tab:summary}. For QPSK at $\rho_1=\SI{6}{\decibel}$ the gate yields $\tau^\star=0.85$ with $10.0\%$, whereas the $\nu=3$ sweep of Fig.~\ref{fig:overlay} gives $\tau^\star=0.75$ with $14.1\%$. The $6$-$\unit{\decibel}$ 16-QAM entry ($3.6\%$ vs.\ $4.5\%$) differs only by the smaller Monte-Carlo budget of this sweep, since the gate does not bind at $\tau^\star=0.90$.

\begin{table}[t]
\centering
\caption{Reliability-gated FTN gain over Nyquist and the corresponding $\tau^\star$ vs.\ reference SNR.}
\label{tab:power}
\begin{tabular}{lccccc}
\toprule
$\rho_1$ [$\unit{\decibel}$] & 0      & 6              & 12             & 18             & 24\\
\midrule
QPSK gain [\%]               & $2.7$  & $10.0^\dagger$ & $24.5^\dagger$ & $25.0^\dagger$ & $25.0^\dagger$\\
QPSK $\tau^\star$            & $0.95$ & $0.85$         & $0.80$         & $0.80$         & $0.80$\\
16-QAM gain [\%]             & ---    & $3.6$          & $7.8^\dagger$  & $21.9^\dagger$ & $25.0^\dagger$\\
16-QAM $\tau^\star$          & ---    & $0.90$         & $0.85$         & $0.80$         & $0.80$\\
\bottomrule
\end{tabular}
\end{table}

\subsection{IAPR Cost of Operating at the Optima}
\label{sec:iapr}
\begin{figure}[t]
\centering
\includegraphics[width=0.92\columnwidth]{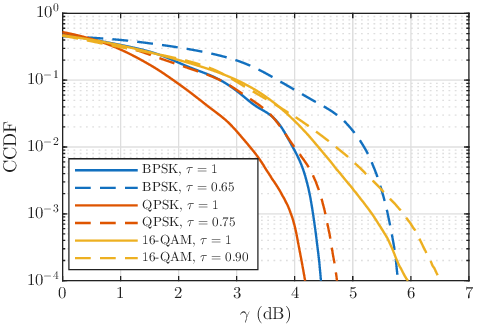}
\caption{IAPR CCDF vs.\ threshold $\gamma$ for BPSK, QPSK, and 16-QAM at $\tau=1$ (solid) and at $\tau=\tau^\star$ (dashed).}
\label{fig:iapr}
\end{figure}

FTN acceleration densifies the pulse overlap, so the transmit waveform $s(t)$ becomes peakier. Fig.~\ref{fig:iapr} shows the
complementary cumulative distribution function (CCDF) of the IAPR, $\Pr(\mathrm{IAPR}>\gamma)$ with $\mathrm{IAPR}=|s(t)|^2/\mathbb{E}[|s(t)|^2]$, at $\tau=1$ and at each SE optimum, for the same RRC pulse as the rate results. Since the normalized IAPR is invariant to power scaling, the curves do not depend on the power convention. At the $10^{-3}$ outage level the acceleration penalty is $\SI{1.3}{\decibel}$ for BPSK ($4.3\!\to\!5.6\,\unit{\decibel}$ at $\tau^\star=0.65$) but only $\SI{0.6}{\decibel}$ for QPSK ($3.9\!\to\!4.5\,\unit{\decibel}$) and 16-QAM ($5.3\!\to\!5.9\,\unit{\decibel}$). As the constellation order grows, the milder optimum also limits the peak-power cost, so the SE gain and the IAPR penalty shrink together. At a common $\tau=0.80$ the ordering is QPSK~$<$~BPSK~$<$~16-QAM, which separates the constellation effect from the acceleration effect. This is consistent with the SNR--IAPR trade-off reported for fixed received power in \cite{zhang2025mimo}, while under the constant transmit-power convention used here the penalty remains bounded.

\section{Conclusion}
Using a common and validated finite-alphabet \emph{symbol-wise} AIR methodology, we showed that FTN acceleration can improve the SE for all considered constellations, but the optimal acceleration becomes less aggressive as the constellation order increases. This trend can be understood through the smaller minimum Euclidean distance of denser constellations under constant-power normalization and its interaction with the ISI introduced by the RRC pulse. A larger-memory 16-QAM detector gives essentially the same rates and the same optimum as the short-memory working detector, so the observed optimum is not a detector artifact. The main trend also persists at lower SNRs. The bit-wise GMI results for coded BICM lead to the same conclusions, with only a small penalty for 16-QAM. The IAPR increase at the FTN operating points remains moderate. Future work will consider coded-system performance and more powerful deep-FTN detectors for higher-order constellations.

\section*{Acknowledgment}
Anthropic's Claude large language model was used for editorial purposes in the preparation of this letter. All methods, results, and conclusions were developed and verified by the authors.

\bibliographystyle{IEEEtran}
\bibliography{refs}

\end{document}